# Securing U.S. Critical Infrastructure: Lessons from Stuxnet and the Ukraine Power Grid Attacks

Jack Vanlyssel

Cybersecurity and National Security

10/8/2024

## Table of Contents

# List of Acronyms and Abbreviations

CISA - Cybersecurity and Infrastructure Security Agency

ICS - Industrial Control Systems

IT - Information Technology

MFA - Multi-Factor Authentication

MS - Microsoft

NIST - National Institute of Standards and Technology

OT - Operational Technology

PLC - Programmable Logic Controller

SANS - SysAdmin, Audit, Network, and Security Institute

SCADA - Supervisory Control and Data Acquisition

SIPROTEC - Siemens Protective Relay (protection system for circuit breakers)

VPN - Virtual Private Network

ZTA - Zero-Trust Architecture

# Introduction

## The National Security Importance of U.S. ICS Cybersecurity

Industrial Control Systems (ICS) form the backbone of critical infrastructure, managing essential services like power grids, water supplies, transportation systems, and manufacturing processes. The U.S. heavily relies on these systems to support both daily life and national security operations [1]. A successful cyberattack on these systems would be catastrophic, disrupting essential services and endangering lives. As more ICS environments integrate with digital infrastructure, their vulnerabilities increase, making them prime targets for cyberattacks with far-reaching national security implications [2]. Ensuring the security of these systems is not only about protecting critical infrastructure but also about safeguarding national stability and resilience against geopolitical threats. The growing sophistication of cyberattacks targeting ICS, demonstrated by the Stuxnet and Ukraine power grid incidents, underscores the urgent need to bolster ICS cybersecurity in the U.S.

## Core Argument

The cybersecurity weaknesses exploited in landmark ICS cyberattacks, such as Stuxnet and the Ukraine power grid attacks, are still prevalent in many U.S. ICS environments, putting the nation's critical infrastructure at severe risk. Despite heightened awareness of these risks, U.S. ICS systems share similar vulnerabilities, including inadequate network segmentation, outdated systems, poor authentication protocols, insufficient monitoring, supply chain weaknesses, and low levels of employee cybersecurity training [3, 4]. If not addressed, these weaknesses could lead to devastating cyberattacks on U.S. infrastructure, causing physical damage, widespread disruptions, and national security crises. Analyzing these vulnerabilities through the lens of past case studies allows the U.S. to learn from past mistakes and enact policies that can mitigate risks and protect our most critical infrastructure from future cyber attacks.

## Scope of the Paper

This paper first examines the fundamental details and critical cybersecurity weaknesses used in the Stuxnet and Ukrainian power grid attacks. Then, the current state of U.S. ICS cybersecurity is analyzed and connected to the vulnerabilities highlighted in the case studies, illustrating how foreign adversaries within the U.S. today can exploit these critical weaknesses. Finally, a policy recommendation aimed at fortifying ICS cybersecurity is offered. This policy addresses network segmentation and zero-trust architecture (ZTA) to secure ICS systems. This paper aims to alert policymakers, industry leaders, and cybersecurity professionals to ICS cyber security weaknesses and suggest policies to safeguard U.S. critical infrastructure against future cyberattacks.

# Stuxnet (2010) – A Turning Point in ICS Cybersecurity

## Background

The 2010 discovery of the Stuxnet by Belarusian cybersecurity firm VirusBlokAda was a watershed moment in the history of ICS cybersecurity. Stuxnet was highly sophisticated malware, considered a "worm" because it replicates itself and spreads from computer to computer without human interaction [5]. It was designed to sabotage Iran's Natanz nuclear facility by targeting programmable logic controllers (PLC), which were ICS that operated uranium enrichment centrifuges [6]. The attack demonstrated the potential of using malware to cripple ICS and critical infrastructure and signaled a new era of cyberwarfare.

## Objectives of the Stuxnet Attack

Widely believed to be a joint effort by the United States and Israel, Stuxnet was designed to covertly disrupt Iran's nuclear capabilities without military intervention. Its primary goal was to sabotage Iran's uranium enrichment process by causing centrifuges to spin at destructive speeds while concealing the damage by showing normal operations to the system's operators [7]. This subtle manipulation made the sabotage appear to be mechanical failure, allowing the malware to avoid detection. Stuxnet sought to buy time for diplomatic nuclear non-proliferation strategies to take effect by delaying Iran's production of weapons-grade uranium. The malware's covert nature allowed attackers to avoid direct military retaliation from Iran while still accomplishing their strategic goals.

# Impact

The Stuxnet attack set back Iran's nuclear program by damaging nearly 1,000 centrifuges, delaying the production of enriched uranium. Strategically, it demonstrated the power of using cyberattacks to inflict physical damage on critical infrastructure, reshaping global perceptions of cybersecurity. Stuxnet's success made nations worldwide reevaluate their ICS security frameworks, encouraging the development of more advanced cybersecurity measures to protect vital infrastructure. The attack also changed the landscape of cyberwarfare and highlighted the vulnerability of ICS across many sectors. As the first publicly known instance of a cyber-physical attack, Stuxnet set a precedent for future cyberattacks that could be used as tools of destruction [8].

# Stuxnet - Key ICS Vulnerabilities

In examining critical vulnerabilities exploited during the Stuxnet attack, we gain valuable insight into how attackers might also target ICS systems in the U.S. Issues such as social engineering, poor network segmentation, supplier vulnerabilities, insufficient intrusion detection, and outdated systems will be covered here and later connected to U.S. cybersecurity. By addressing similar vulnerabilities, the U.S. can better defend against potential attacks on its ICS environments, which could otherwise suffer similar consequences as those inflicted by Stuxnet.

## Social Engineering

The introduction of Stuxnet into the Iranian Industrial Control Systems (ICS) network at the Natanz nuclear facility was only possible through social engineering tactics. U.S. and Israeli intelligence likely had intelligence agents posing as contractors give infected USBs to employees at the Iranian facility. These employees then infected the entire network by plugging the USBs into the air-gapped computers at the facility (air-gapped, meaning isolated from external networks, such as the internet, for security purposes) [9]. This social engineering attack exploited human trust, tricking employees into introducing malicious software into the system by relying on their lack of caution or awareness about potential security threats. Without using this technique, it is unlikely the Natanz facility would have been infected at all.

## Lack of Network Segmentation

Once inside the Natanz facility, Stuxnet exploited poor network segmentation to spread across critical systems. Iran relied on air-gapping for security and assumed it would be enough to protect ICS systems completely. The lack of internal network segmentation allowed Stuxnet to access the entire network after being introduced to one system [10]. Proper network segmentation would have limited the spread of Stuxnet and required each system to be infected individually. This would allow more time for the malware to be detected and would have saved many ICS. While air-gapping is an effective protective measure, it can create a false sense of security based on the assumption that attackers will never reach critical networks, which was exploited in this case by Stuxnet.

## Supply Chain Vulnerabilities

Stuxnet took advantage of many supply chain vulnerabilities to compromise Natanz ICS. Firstly, Stuxnet used four unique zero-day exploits in the Microsoft Windows operating system to spread stealthily and infect systems without detection. These zero-day exploits were previously unknown software vulnerabilities that attackers exploited before the software vendor became aware and issued a patch, making it highly dangerous. Second, Stuxnet used stolen digital certificates from Realtek and JMicron to make its malicious code appear as legitimate software. This allowed Stuxnet to bypass security measures and execute its payload. Finally, once inside a targeted environment, Stuxnet exploited a vulnerability in Siemens Step7 software, which operated the PLC at Natanz. Through this vulnerability, Stuxnet injected code into the PLCs without alerting operators [6]. From propagation to the malicious code used to destroy centrifuges, the entire Stuxnet attack was enabled by supply chain vulnerabilities, which underscores the need for more robust security in vendor selection.

## Insufficient Intrusion Detection

A fundamental weakness during the Stuxnet attack was the absence of effective intrusion detection and response systems within Iran's ICS networks. Stuxnet silently infiltrated and spread for months because it transmitted false data to centrifuge operators, making it appear that centrifuges were working as normal when, in reality, the facility's ICS had been compromised. In addition, while Iran had Information Technology (IT) monitoring systems in place, they lacked the necessary depth of monitoring to detect anomalies in digital traffic caused by Stuxnet. Without sufficient monitoring or tools for cross-verifying sensor data against real-world operational behavior, Stuxnet was able to spread and wreak havoc, destroying around 1000 centrifuges. Iran could have prevented a lot of physical damage to the Natanz facility if it had employed strong IT and ICS-specific intrusion detection tools [11].

## Old and Unpatched Systems

While this issue is not specific to Iran, it is essential to note that Stuxnet exploited the reliance on outdated and unpatched systems within the Natanz ICS infrastructure. The facility used legacy systems that were particularly vulnerable to zero-day attacks. These older systems lacked updates and security patches to defend against sophisticated malware like Stuxnet. Some of the zero-day exploits used by Stuxnet, such as MS10-061, had been patched, but the facility's reliance on outdated technology prevented these patches from being applied effectively [6]. Many ICS environments globally, especially within critical infrastructure, still depend on legacy systems that are not regularly updated. The use of unpatched systems poses significant security risks, as these environments often cannot apply new patches or updates in real time due to operational constraints. Failing to address vulnerabilities in older systems leaves facilities open to similar cyberattacks.

# Ukrainian Power Grid Cyber Attacks (2015 & 2016) – ICS Warfare

## Background and Context

The Ukrainian cyberattacks in 2015 and 2016 were a pivotal moment in the history of ICS cybersecurity, marking the first known successful attacks to disrupt a nation's power grid. In December 2015, a coordinated attack using the BlackEnergy 3 malware targeted three energy distribution companies' Supervisory Control and Data Acquisition (SCADA) systems. SCADA is an ICS that monitors and controls critical infrastructure like power grids, allowing attackers to manipulate circuit breakers remotely, cutting power to approximately 225,000 people. The following year, a more sophisticated cyberattack used the "Industroyer" malware to target a transmission substation's ICS, leading to further outages. Both incidents were attributed to Russian state-sponsored actors, occurring amidst heightened geopolitical tensions following Russia's annexation of Crimea. These attacks demonstrated new vulnerabilities in ICS systems and were an escalation in the use of cyber capabilities against civilian infrastructure [12].

## Objectives of the Ukraine Attacks

Russia's primary objective was to disrupt civilian life on a large scale and demonstrate its ability to take down critical infrastructure using cyber attacks. The broader goal was to instill fear in people and undermine public confidence in the Ukrainian government's ability to protect its infrastructure during their ongoing conflict. Politically, the attacks aimed to show Europe and the rest of the world that their critical infrastructure systems could be similarly attacked if they were to engage in conflict with Russia. Additionally, the attacks allowed Russia to test advanced cyber espionage strategies for future use against its adversaries [13].

## Impact

The immediate impact of the attacks was power outages affecting hundreds of thousands of people, which created widespread fear as Russia had intended. The attacks exposed Ukraine's lack of preparedness against sophisticated cyber threats. This, combined with the invasion of Crimea a year earlier, led to a loss of trust and confidence in the government. Internationally, the attacks heightened concerns about the vulnerability of critical infrastructure, particularly in Europe and the U.S. Compared to Stuxnet, the Ukrainian power grid cyberattacks were far more aggressive and overt, breaking new ground by explicitly using cyber tools to create fear and chaos, setting a new precedent for weaponizing cyber capabilities in modern warfare and exposing severe vulnerabilities in ICS cybersecurity globally [14].

# Ukraine - Key ICS Vulnerabilities

Many of the vulnerabilities exploited in the 2015 and 2016 Ukrainian power grid attacks share commonalities with those seen in the Stuxnet attack, including weak authentication protocols, poor network segmentation, insufficient intrusion detection, and outdated systems. These overlapping vulnerabilities highlight the urgent need to address similar weaknesses in U.S. ICS environments. Despite these common themes, the Ukrainian attacks offer new lessons, particularly as the first successful cyberattack on a power grid. The U.S. can gain crucial insights and significantly improve its ICS cybersecurity using lessons from this incident.

## Phishing and Social Engineering

Both the 2015 and 2016 attacks on Ukraine's power grid began with targeted "spear phishing" emails sent to employees of various Ukrainian energy companies. These emails contained malicious Microsoft Excel and Word files, which, once opened, used macros to deploy malware to Ukrainian systems [15]. Macros are automated scripts embedded within documents that can execute a series of commands, and when enabled, they can be exploited by attackers to run malicious code. This method allowed attackers to initially gain access to IT networks, from which they moved laterally to compromise ICS systems controlling critical infrastructure. Similar to the Stuxnet case, a lack of employee training was responsible for the initial infection of Ukrainian systems.

## Weak Authentication & Remote Access Security

During the 2015 incidents, attackers gained access to the SCADA systems using virtual private network (VPN) credentials stolen using the BlackEnergy 3 malware that was spread in the aforementioned phishing attack. There were no multi-factor authentication systems in place, so attackers could freely use stolen accounts to appear as legitimate users, allowing them to control Ukrainian ICS systems remotely. In the 2016 attacks, the malware Industroyer took advantage of communication protocols IEC 60870-5-101 and IEC 61850 to remotely access SCADA systems to disable circuit breakers, shutting off power to Kyiv for just over an hour [16]. An ever-increasing number of ICS systems are being connected to the internet today for remote use. This is especially true for the electric sector, which has many geographically dispersed systems that can be difficult to coordinate manually. Remote access can increase the ease of use of ICS systems for both employees and bad actors if not properly secure.

## Lack of Network Segmentation

In both the 2015 and 2016 power grid hacks, improper network segmentation between information technology and operational technology systems was taken advantage of. Once inside the networks, Russian hackers were able to escalate their privileges and move laterally across the network from the IT network to the ICS that was running power grid operations. There was no separation between the corporate networks and operations technology. Once attackers infiltrated part of the network, they had access to all of it [17]. Better network segmentation could have significantly limited attackers' ability to move across the network, restricting their access to the ICS that operated the grid.

## Insufficient Intrusion Detection and Response

Once the attackers had access to the internal network of the Ukrainian power grid, they were able to remain in the system for months, preparing for the attack without setting off any alarms. This indicates a lack of monitoring and detection within the network. In addition, the response to the attack was manual and reactive, relying on operators physically restoring power by visiting substations and resetting systems manually. This delayed recovery indicated that there were no incident response protocols in place. Incorporating more advanced detection systems could have stopped the reconnaissance of attackers, preventing them from learning enough to conduct such a coordinated attack [17].

## Old and Unpatched Systems

Similar to Iran, the Ukrainian power grid relied on outdated and unpatched systems, which were critical in attackers' ability to shut off power. Many of the control systems used in these attacks had known vulnerabilities that were not addressed due to a lack of regular updates. In the 2016 Ukrainian power grid attack, the Industroyer malware specifically targeted ICS controlling SIPROTEC protective relays, which are designed to open circuit breakers when detecting dangerous electrical conditions [18]. The attackers exploited a flaw that allowed a single malicious network packet to disable these relays, rendering them useless until manually rebooted. Siemens issued a patch in 2015 to fix the vulnerability, but many relays in Ukraine were not updated, leaving them vulnerable. Failure to update software allowed attackers to target documented vulnerabilities, enabling them to gain control of SCADA systems. A more robust patch management strategy would strengthen security posture and may have prevented at least one of the attacks.

# U.S. ICS Vulnerabilities: Lessons from Case Studies

The cybersecurity weaknesses exploited in the case studies are not isolated to Ukraine and Iran. The same vulnerabilities are widespread in U.S. industrial control systems as well, despite a growing awareness of cyber threats to critical infrastructure. By comparing weaknesses in ICS security attackers exploited in Stuxnet and Ukraine to the current state of security in the U.S., we can learn valuable lessons about what vulnerabilities are present in America's ICS cybersecurity.

## Social Engineering and Employee Training

In the Stuxnet attack, social engineering played a critical role in introducing the malware into Iran's isolated ICS network. Attackers used infected USB drives, which were either deliberately planted or unknowingly introduced by employees and then used by unknowing system operators. Similarly, the 2015 Ukraine power grid attack was initiated through spear-phishing emails that tricked employees into downloading malicious attachments, granting attackers access to the network. As demonstrated by these attacks, social engineering poses a significant vulnerability in ICS environments where operators have little training. In the U.S., a 2021 National Cybersecurity Alliance report found that 40% of ICS employees had received little to no formal cybersecurity training, increasing the risk of exploitation through human error [19]. Social engineering is the easiest way for attackers to infect target systems, and adversaries will continue to use social engineering tactics well into the future. Without proper training, the U.S. is susceptible to similar tactics used in Iran and Ukraine to infect ICS systems with destructive malware.

## Lack of Network Segmentation

A significant vulnerability exploited in both the Stuxnet and Ukraine attacks was the lack of proper network segmentation. In Stuxnet, even though the Iranian nuclear facility was air-gapped, the malware was able to spread internally due to insufficient segmentation between different control systems. Similarly, in Ukraine, attackers moved laterally through the power grid network after gaining access via phishing emails, highlighting the absence of adequate barriers between IT and operational technology (OT) systems. This issue is prevalent in many U.S. ICS environments as well, where IT and OT networks are often interconnected without stringent controls. According to a "SysAdmin, Audit, Network, and Security Institute" (SANS) survey on U.S. ICS cybersecurity, more than 60% of respondents reported that their ICS environments were connected to external networks, and nearly half acknowledged a lack of adequate network segmentation [20]. Without better network isolation, malware or unauthorized users can move freely between critical systems, increasing the risk of widespread disruption.

## Insufficient Authentication & Access Controls

The poor authentication and access control mechanisms exploited in the Ukraine power grid attacks are also significant weaknesses in U.S. ICS environments. In Ukraine, attackers gained remote access to the SCADA systems using stolen VPN credentials, facilitated by the absence of multi-factor authentication (MFA) and weak password policies. U.S. ICS systems face similar challenges, as many critical infrastructure operators still rely on basic authentication methods, and many systems lack MFA entirely. The Cybersecurity and Infrastructure Security Agency has repeatedly warned that weak authentication practices in ICS networks remain a significant vulnerability. A Cybersecurity and Infrastructure Security Agency (CISA) report from 2021 highlighted numerous instances where attackers gained unauthorized access to ICS systems in the U.S. by exploiting weak or default passwords. One example is the 2021 Oldsmar, Florida water treatment plant attack, where attackers gained access to a SCADA system through TeamView ( which did not have MFA set up) and attempted to pollute the town's water supply, showing that this remains an ongoing risk in American critical infrastructure [21].

## Poor Incident Detection and Response

Both the Stuxnet and Ukraine power grid attacks demonstrated the consequences of poor incident detection and response capabilities. In both cases, the malware went undetected for long periods, allowing attackers to operate within the network without triggering alarms. This same weakness is present in many U.S. ICS systems, where the lack of specialized intrusion detection systems for ICS environments leaves critical infrastructure vulnerable to stealthy attacks. A 2022 Dragos report on U.S. ICS cybersecurity found that over 70% of ICS environments lacked dedicated monitoring systems for OT networks, making it difficult to detect or respond to intrusions in real-time [22]. This gap in monitoring and response is difficult, given the increasing sophistication of cyberattacks targeting U.S. infrastructure.

## Supply Chain Vulnerabilities

The Stuxnet attack underscored the risk of supply chain vulnerabilities in protecting ICS. The worm spread using Microsoft Windows exploits and destroys centrifuges by exploiting vulnerabilities in Siemens PLC software. The attack highlighted how third-party suppliers can introduce critical weaknesses into ICS environments. The U.S. faces similar risks, as ICS environments heavily depend on third-party vendors for both hardware and software. The 2020

SolarWinds attack demonstrated that attackers could infiltrate American critical infrastructure by compromising trusted software suppliers. In this case, hackers inserted malicious code into updates for SolarWinds' Orion software, which was widely used by government agencies and private companies. The compromised updates allowed attackers to gain undetected access to systems for months, causing widespread damage [23]. In addition, supply chain vulnerabilities remain a significant concern because many American ICS use outdated or untested third-party components, which may have hidden vulnerabilities or backdoors.

## Outdated and Unpatched Systems

Stuxnet was able to use zero-day exploits due to outdated and unpatched systems in the Iranian nuclear facility. In Ukraine, too, many of the ICS devices, including the SCADA systems, were running on old software with known vulnerabilities that had not been patched, allowing attackers to gain access and cause widespread power outages. In the U.S., many ICS systems, particularly in critical sectors like energy, water, and manufacturing, still rely on legacy hardware and software that are difficult to update without interrupting operations. A 2020 Ponemon Institute report found that 56% of U.S. organizations using ICS had experienced cyber incidents in the past two years, and many of these incidents were attributed to unpatched vulnerabilities in outdated systems [24]. The complexity and cost of updating ICS environments lead to delays in patching known vulnerabilities, leaving critical infrastructure at risk of attackers exploiting techniques similar to Stuxnet and Ukraine.

# Preventing the Exploitation of Weaknesses in U.S. ICS

With the vulnerabilities exploited in the Stuxnet and Ukraine power grid attacks now confirmed in U.S. ICS environments, the extent of the risks is clear. Strengthening cybersecurity practices in ICS environments will require a combination of technical measures, new policies, and regulatory frameworks, but all of the exploited weaknesses can be addressed through targeted solutions. This section will detail all of the possible approaches to strengthening security so that a policy recommendation can be made.

## Social Engineering & Employee Training

Due to the demonstrated issues with social engineering, U.S. ICS environments must implement a comprehensive approach that addresses human vulnerabilities. In both case studies, attackers exploited untrained employees, gaining access to secure networks through phishing emails and infected USB drives. To mitigate these risks, organizations should prioritize cybersecurity training for all employees, focusing on recognizing phishing attempts, safely handling suspicious emails, and understanding the dangers of removable media like USB drives. Additionally, continuous phishing simulations can reinforce training by testing employees' responses to simulated attacks, identifying those vulnerable to manipulation, and improving awareness. These proactive measures, combining technical controls with human-focused strategies, would significantly reduce the risk of initial exposure of ICS to malware.

## Lack of Network Segmentation

Network segmentation must be strictly enforced to prevent malware from spreading freely within ICS environments. The National Institute of Standards and Technology recommends implementing network zones with firewalls (to filter traffic) and demilitarized zones (to create buffers) between IT and OT environments. Additionally, network segmentation technologies like Virtual Local Area Networks and intrusion prevention systems tailored to OT environments are recommended to help detect and block lateral movements within the network enforcing segmentation [25]. Furthermore, zero-trust architecture is gaining popularity to ensure that every device, user, and network segment is treated as potentially compromised. By implementing strict authentication and authorization policies for all internal traffic between segmented networks, attackers would face significant challenges in moving laterally, even after gaining initial access. These restrictive segmentation policies can significantly reduce the chances of malware spreading within critical systems.

## Insufficient Authentication and Access Controls

Based on the discussed vulnerabilities, multi-factor authentication must be universally implemented in U.S. ICS environments. Attackers in the Ukraine incidents could gain remote access to SCADA systems using stolen VPN credentials, which were not protected by MFA. In U.S. ICS systems, many operators still rely on basic authentication methods, leaving critical infrastructure vulnerable to similar exploits. CISA recommends enforcing MFA for all users, particularly for remote access points like VPNs and SCADA systems [26]. Another essential step is to improve password management by eliminating default or weak credentials that attackers can easily exploit. U.S. ICS systems should adopt strong password policies that mandate regular updates, complexity, and unique credentials for each user. These updates would prevent U.S. ICSs from being exploited like Ukrainian ICSs were.

## Poor Incident Detection & Response

In both the Stuxnet and the Ukraine attacks, hackers could operate within the network for extended periods without detection, causing significant damage. To prevent similar scenarios, U.S. ICS environments must deploy dedicated Intrusion Detection Systems explicitly designed for operational technology networks, such as the Dragos Platform, which offers real-time monitoring, threat detection, and asset visibility tailored to OT traffic [27]. Implementing Dragos or a similar system would allow for real-time monitoring of ICS traffic and help detect anomalies before significant damage occurs. Additionally, a proactive incident response plan should be in place to swiftly react to detected intrusions. The manual, reactive response in Ukraine, where operators had to physically visit substations to restore power, highlights the need for automated, centralized control and response mechanisms to mitigate attacks without delay.

## Supply Chain Vulnerabilities

Supply chain vulnerabilities played a significant role in both case studies and have been exploited numerous times in America. While difficult to secure, rigorous supply chain vetting and monitoring are necessary to ensure ICS security. U.S. ICS owners should only work with trusted suppliers who follow strong cybersecurity standards and implement secure software development lifecycles. Additionally, software and hardware vendors should be required to regularly update and patch vulnerabilities in their products to prevent exploitation. Vendor penetration testing and security audits would also further secure the supply chain. Addressing these supply chain weaknesses could significantly reduce vulnerability to the exploits seen in Stuxnet and Ukraine.

## Outdated and Unpatched Systems

Iran's use of legacy systems with unpatched vulnerabilities allowed the worm to infiltrate and disrupt critical infrastructure. Similarly, the Ukraine attacks capitalized on unpatched SCADA systems to manipulate the power grid. U.S. ICS environments must prioritize patch management and regular system updates to mitigate similar risks. Organizations should implement automated tools that regularly check for updates and apply security patches to ICS and OT systems.

# Policy Remediation: Zero-Trust Network Segmentation

From the case studies, it's clear that ICS remain vulnerable to cyberattacks that exploit poor network segmentation. In both cases, malware spread across multiple systems due to weak segmentation between Information Technology and Operational Technology networks. A policy solution to this issue is mandating the implementation of zero-trust architecture and stricter network segmentation. This policy would significantly reduce the risk of lateral movement by attackers within U.S. ICS environments, thereby protecting critical infrastructure from both internal and external threats [25].

## Core Components

This policy would utilize micro-segmentation, breaking down networks into smaller, manageable segments. Each segment would be isolated and treated as an independent entity, with access control measures applied at every level. This would ensure that if an attacker gains access to one portion of an ICS, they cannot easily move to other parts of the network. Within these micro-segments, continuous authentication and authorization would require all users, devices, and systems to continuously prove their identity and authorization status before being granted access to any network segment. In addition to this constant authorization, users and devices would be given the minimum level of access necessary to perform their functions to limit damage from attackers gaining access to a system. Finally, this new system would feature robust logging and monitoring throughout the ICS environment. When users or devices attempt to access micro-segments, the ICS will log the attempt [28]. These logs allow for early detection of suspicious activity and rapid response before significant damage occurs.

## Why This Policy Is Best

The policy of enforcing network segmentation through zero-trust architecture offers the most comprehensive and effective means of securing ICS environments. Traditional perimeter-based defenses must be improved in the face of modern, sophisticated cyberattacks that can bypass external defenses through supply chains, social engineering, and outdated technology exploits. The key strength of this policy lies in its assumption that every part of the network can and will be compromised, requiring continuous validation of all entities within the network. This approach directly addresses the weaknesses observed in the Stuxnet and Ukraine attacks, where poor network segmentation allowed attackers to move laterally within the network and target critical systems. This policy ensures that even if attackers were to gain access, the micro-segmentation and least privilege access controls would prevent them from moving, causing widespread damage. By applying strict access controls, continuous verification, and robust monitoring, ZTA provides a multi-layered defense that significantly reduces the risk of a catastrophic breach in U.S. ICS systems.

## Feasibility

While this policy represents a significant shift in cybersecurity practices, it can be incrementally implemented, allowing organizations to start with high-priority segments of their ICS environments. This step-by-step approach reduces the complexity of a full-scale implementation and allows for gradual adoption without disrupting operations. In addition, many core technologies required for ZTA, such as micro-segmentation, continuous monitoring, and identity management systems, are already mature and widely available and can be integrated with existing ICS infrastructure, avoiding costly overhauls. Organizations such as the National

Institute of Standards and Technology and the Cybersecurity and Infrastructure Security Agency have already issued guidelines that support the principles of ZTA [29]. These guidelines provide a framework for organizations to follow, ensuring that they can implement ZTA in a standardized and compliant manner. The initial costs of implementing ZTA may be high due to the need for new technologies and training. Still, the long-term benefits of preventing a large-scale cyberattack outweigh these upfront investments. In terms of financial losses and damage to national security, the cost of recovering from an attack like Stuxnet or the Ukraine grid hack is far greater than securing ICS systems through ZTA.

# Conclusion

The cyberattacks on industrial control systems, exemplified by the Stuxnet and Ukraine power grid incidents, serve as stark warnings about the vulnerabilities in critical infrastructure systems. These attacks revealed significant weaknesses in network segmentation, outdated and unpatched systems, weak authentication and access controls, poor incident detection and response, supply chain vulnerabilities, and low levels of employee cybersecurity training. As demonstrated, these same weaknesses are prevalent in U.S. ICS environments, leaving the nation's essential services susceptible to similar cyberattacks. Protecting the U.S. critical infrastructure requires learning from these past incidents and implementing robust cybersecurity measures that address technological and human factors.

## Key Lessons

The lessons from the Stuxnet and Ukraine attacks highlight the need for proactive measures in ICS cybersecurity. First, network segmentation is critical to preventing malware from moving within systems. Second, patch management and system modernization must be prioritized to address the vulnerabilities inherent in outdated systems. Third, more robust authentication protocols are necessary to prevent unauthorized access, especially in environments where remote access is essential. Fourth, improved detection systems are needed to identify and respond to threats, minimizing potential damage quickly. Fifth, supply chain vulnerabilities must be addressed through stringent security practices for third-party vendors. Finally, human factors, such as employee awareness and training, are equally as important as technical measures for ensuring cybersecurity.

## Final Recommendations

The most critical step to fortifying the security of U.S. Industrial Control Systems is the implementation of zero-trust architecture combined with strict network segmentation. These measures should be mandated for all ICS environments, particularly those managing critical infrastructure such as energy, water, and transportation systems. By segmenting networks into smaller, isolated units and ensuring that every device, user, and system is continuously authenticated and authorized, this policy will prevent the lateral movement of attackers within ICS networks. Deploying real-time monitoring and logging systems further enhances the ability to detect and respond to potential threats. At the same time, a least privilege access model limits the damage any unauthorized access can cause. This comprehensive approach addresses the core vulnerabilities exploited in the Stuxnet and Ukraine attacks, where weak network segmentation allowed attackers to wreak havoc across ICS systems. By adopting ZTA and robust network segmentation, U.S. critical infrastructure will have a multi-layered defense that significantly mitigates the risk of large-scale cyber attacks.

## Call to Action

U.S. policymakers, industry leaders, and cybersecurity experts must prioritize the standardization of zero-trust architecture and network segmentation within ICS environments. The lessons from the Stuxnet and Ukraine power grid attacks provide a clear blueprint for the dangers of inadequate cybersecurity measures. A coordinated national effort is required to mandate these security upgrades, ensuring that technical and human vulnerabilities are addressed. Enacting this policy will protect the nation's critical infrastructure and ensure its resilience in the face of increasingly sophisticated cyber threats.

# Sources

[1] - *Critical Infrastructure Sectors: CISA*. (n.d.). Cybersecurity and Infrastructure Security Agency CISA.
https://www.cisa.gov/topics/critical-infrastructure-security-and-resilience/critical-infrastructure-sectors

[2] - National Cybersecurity and Communications Integration Center (NCCIC) Industrial Control Systems Cyber Emergency Response Team (ICS-CERT). (2016). Recommended practice: Improving industrial control system cybersecurity with defense-in-depth strategies. U.S. Department of Homeland Security.
https://www.cisa.gov/sites/default/files/recommended_practices/NCCIC_ICS-CERT_Defense_in_Depth_2016_S508C.pdf

[3] - Lendvay, R. L. (2016). *Shadows of Stuxnet: Recommendations for U.S. policy on critical infrastructure cyber defense derived from the Stuxnet attack*.
https://www.hsaj.org/articles/10937

[4] - Shehod, A. (2016). Ukraine Power Grid Cyberattack and US Susceptibility: Cybersecurity Implications of Smart Grid Advancements in the US. *Sustainable Energy*.
https://web.mit.edu/smadnick/www/wp/2016-22.pdf

[5] - Zetter, K., & Ochman, J. (2014). *Countdown to Zero Day: Stuxnet and the launch of the world's first Digital Weapon*. Books on Tape.

[6] - Symantec. (2011). W32.Stuxnet dossier. Symantec Security Response.
https://www.symantec.com/content/en/us/enterprise/media/security_response/whitepapers/w32_stuxnet_dossier.pdf

[7] - Lindsay, J. R. (2013). Stuxnet and the limits of cyber warfare. Security Studies, 22(3), 365-404.
https://doi.org/10.1080/09636412.2013.816122

[8] - Farwell, J. P., & Rohozinski, R. (2011). Stuxnet and the future of cyber war. Survival, 53(1), 23-40.
https://doi.org/10.1080/00396338.2011.555586

[9] - *Stuxnet explained: The first known cyberweapon*. CSO Online. (2022).
https://www.csoonline.com/article/562691/stuxnet-explained-the-first-known-cyberweapon.html

[10] - Armis Security. (2021). *Chapter 7 – Network Segmentation: A Cybersecurity Best Practice to Protect Industrial Assets*.
https://www.armis.com/blog/chapter-7-network-segmentation-a-cybersecurity-best-practice-to-protect-industrial-assets

[11] - Storm, J.-M., Houmb, S. H., Kaliyar, P., & Erdodi, L. (2024). Testing Commercial Intrusion Detection Systems for Industrial Control Systems in a Substation Hardware in the Loop Testlab. Electronics, 13(1), 60.
https://doi.org/10.3390/electronics13010060

[12] - Wingreen, S. C., & Samandari, A. (Eds.). (2024). *Information technology security and risk management: Inductive cases for information security* (1st ed.). CRC Press.
https://doi.org/10.1201/9781003264415

[13] - U.S. Department of Justice. (2020). *Six Russian GRU officers charged in connection with worldwide deployment of destructive malware and other disruptive actions in cyberspace*.
https://www.justice.gov/opa/pr/six-russian-gru-officers-charged-connection-worldwide-deployment-destructive-malware-and

[14] - Malekos Smith, Z. L. (2022). A power struggle over Ukraine's electrical grid. Center for Strategic and International Studies.
https://www.csis.org/analysis/power-struggle-over-ukraines-electrical-grid

[15] - Beek, C. (2016). A case of mistaken identity? The role of BlackEnergy in Ukrainian power grid disruption. McAfee.
https://www.mcafee.com/blogs/other-blogs/mcafee-labs/blackenergy_ukrainian_power_grid/

[16] - CISA. (2016). *Cyber-attack against Ukrainian critical infrastructure*.
https://www.cisa.gov/news-events/ics-alerts/ir-alert-h-16-056-01

[17] - SANS Institute. (2016). *Confirmation of a coordinated attack on the Ukrainian power grid*.
https://www.sans.org/blog/confirmation-of-a-coordinated-attack-on-the-ukrainian-power-grid/

[18] - Greenberg, A. (2019). *New clues show how Russia's grid hackers aimed for physical destruction*. WIRED.
https://www.wired.com/story/russia-ukraine-cyberattack-power-grid-blackout-destruction/

[19] - National Cybersecurity Alliance. (2021). *2021 Cybersecurity workforce study*.
https://iapp.org/resources/article/isc2-2021-cybersecurity-workforce-study

[20] - *Kim, A*. SANS Institute. (2024) Five Startling Findings In 2023's ICS Cybersecurity Data.
https://www.sans.org/blog/five-startling-findings-2023-ics-cybersecurity-data/

[21] - Cybersecurity and Infrastructure Security Agency. (2022). *Understanding and mitigating Russian state-sponsored cyber threats to U.S. critical infrastructure (Advisory AA22-011A)*. U.S. Department of Homeland Security.
 https://www.cisa.gov/news-events/cybersecurity-advisories/aa22-011a

[22] - Dragos, Inc. (2023). *2022 Year in Review: Industrial Control Systems & Operational Technology Cybersecurity Report*.
https://www.dragos.com/blog/industry-news/2022-dragos-year-in-review-now-available/

[23] - CISA. (2020). *Alert (AA20-352A) - Advanced Persistent Threat Compromise of Government Agencies, Critical Infrastructure, and Private Sector Organizations*. Cybersecurity and Infrastructure Security Agency.
https://us-cert.cisa.gov/ncas/alerts/aa20-352a

[24] - Ponemon Institute. (2021). *The 2021 state of industrial cybersecurity*. Dragos.
https://www.dragos.com/resources/reports/2021-state-of-industrial-cybersecurity-ponemon

[25] - Cybersecurity and Infrastructure Security Agency. (2020). Securing industrial control systems. U.S. Department of Homeland Security.
https://www.cisa.gov/resources-tools/resources/securing-industrial-control-systems

[26] - Cybersecurity and Infrastructure Security Agency. (2022). Control system defense: Know the opponent (AA22-265A). U.S. Department of Homeland Security.
https://www.cisa.gov/news-events/cybersecurity-advisories/aa22-265a

[27] - National Institute of Standards and Technology. (2022). *Guide to operational technology (OT) security* (NIST Special Publication 800-82r3). U.S. Department of Commerce.
https://doi.org/10.6028/NIST.SP.800-82r3

[28] - National Institute of Standards and Technology. (2020). *Zero trust architecture* (NIST Special Publication 800-207). U.S. Department of Commerce.
https://doi.org/10.6028/NIST.SP.800-207

[29] - Cybersecurity and Infrastructure Security Agency. (2021). *Zero Trust Maturity Model*. U.S. Department of Homeland Security.
https://www.cisa.gov/zero-trust-maturity-model